# Kannudi - A Reference Editor for Kannada

(Based on OPOK! and OHOK! principles, and Domain Knowledge)

Vishweshwar V. Dixit

KRIA, Kannada Research Institute of America

714-322-9748 namovish@kannadakali.com; namovish@gmail.com

## Abstract

Kannudi is a reference editor for Kannada based on OPOK! and OHOK! principles, and domain knowledge. It introduces a method of input for Kannada, called OHOK!, that is, *Ottu Hāku Ottu Koḍu!* (apply pressure and give ottu). This is especially suited for pressure sensitive input devices, though the current online implementation uses the regular mechanical keyboard. OHOK! has three possible modes, namely, sva-ottu (self-conjunct), *kandante* (as you see), and *andante* (as you say). It may be noted that *kandante* mode does not follow the phonetic order. However, this mode may work well for those who are inclined to visualize as they type rather than vocalizing the sounds.

Kannudi also demonstrates how domain knowledge can be effectively used to potentially increase speed, accuracy, and user friendliness. For example, selection of a default vowel, automatic *shunyification*, and *arkification*. Also implemented are four types Deletes that are necessary for phono-syllabic languages like Kannada.

Kannudi can be accessed at
https://kannadakali.com/kannudi/kannudi.html

## 1 Introduction

Many tools are available for digital inputting of Kannada and other Indian language text, such as Input Method Editors (IME), and real time transliteration tools, free and commercial, online as well as offline. Most of these are generic in the sense that they are designed to address all Indian languages. That makes sense as the scripts for these languages, as they all descended from the same Brahmi script, share many common features such as being alpha-syllabic. However, there are subtle differences in the writing styles of these languages. These language specifics can be used to "optimize" and make the editors and IMEs more efficient and user friendly. An earliest implementation of one such editor attempting to use the 'domain knowledge' was described by Dixit [1] [2] [3]. It was an ambitious effort as it aimed for a universal framework. It identified notions such as *Śūn'yīfication*, *Non-initial Vowel*, and hinted at *a-rule* for voice input. However, the implementation was limited to Kannada and limited in scope of the features. This was a DOS based editor and no versions were released later on Windows or other platforms. Another notable DOS editor was developed in Visual Basic by Rangachar [4].

Described in this article is a reference implementation introducing a new method of input for *ottakshara*s (conjuncts) while incorporating previous ideas in Chitragupta and Unived [1] [2].

## 2 Input Methods

**Keyboarding,** being the norm, is a required method of input. A phonetic input method generally assigns a single key to a single phoneme and the keys are typed in the order of pronunciation. A strict one phoneme - one key (OPOK!) mapping of keys to phonemes may not desirable for the sake of convenience. Some may assign multiple combinations of 1-3 keys to a single phoneme for convenience to resemble common writing using Latin script. For example, one may assign *B, bh,*

*Bh*, or *BH* for ಭ (*mahāprāṇa b*) for the sake of convenience.

A mapping of letters to (ASCII) keys was developed by Kannada Ganaka Parishattu (KGP) and is used in its Nudi editor [5]. This has been designated as the official standard by Government of Karnataka [6].

A context sensitive dynamic keyboarding has been described by Joshi *et.al*. [7]

**Handwriting**, using stylus on a tablet or mobile screen, is another method of input. The complexity of graphics processing makes it slow. Variations in individual writing styles also introduces errors. Correcting these errors as one goes takes time and interrupts the thought process of the user which reduces the speed.

**Voice** input is a promising method. However, Voice recognition is not perfect. This is especially problematic in Kannada where regional and other variations in pronunciations abound. Additionally, current implementations are mostly dictionary based. They suffer from the same difficulties with all Indian languages, namely, variations in pronunciation, dictionary limitations, and indetermination between writing as pronounced and the dictionary entries.

Current voice input implementations may make it more difficult to type. One needs to keep looking constantly at the words being entered, select among the choices, or correct manually. A user seems to spend significant time in 'correcting' the dictionary words and ultimately, being frustrated, ends up using or taking help of a basic keyboard layout such as Inscript. Thus, voice input, though convenient, current implementations fail the user in both accuracy and speed.

## 3 Improvement Opportunities

Three main considerations in any input method are accuracy, speed, and convenience. Appropriate tradeoff among the three is also of concern. Elimination or minimization of corrections (backspace and deletes) and minimization of required keystrokes become important parameters. Certainly, there is a need and room to improve the existing methods in these regards.

Current implementations aim and try to cater to all Indian languages and therein implementing a set of common minimum features. They become minimally useful, as most users are not interested in 15+ languages. Hence, an implementation must be extensible so that domain knowledge specific to each language and script can be incorporated

Mobile platforms present interesting opportunities for novel methods using soft keyboards, dynamic context, and new input mechanisms such as swipes.

## 4 Letter Frequencies

Whereas earlier studies have found 36% *mūla akṣaras* (vowel *a* or *C+a*) and 14% conjuncts, using a sample of articles from Kannada Wikipedia [8] and Kannada Kali [9], we found the letter frequencies as shown in Table 1.

| *moola akshara* | 42.4% |
| --- | --- |
| *Gunitakshara* | 39.3% |
| anusvāra / sonne / śūn'ya | 5.6% |
| *Ottu and end virāma* | 18.2% |
| *Sajāti/Dvitva (self-conjunct)* | 8.7% |
| *Dvitva post-vowel* | 1.1% |
| *Vijāti (non-self-conjunct)* | 8.8% |
| *End-virāma* | 0.7% |

Table 1: Syllable Frequencies

## 5 Kannudi

Kannudi [10] is a reference editor introducing several innovative and experimental features. Currently, an online implementation invoked via a web browser, is available. Some features in this implementation require a keyboard. Kannudi input method follows the phonetic order, i.e., phonemes are entered in the order of their pronunciation.

Major principles in Kannudi are OPOK!, OHOK!, user friendliness, prevention/minimization of errors and illegal combinations and letter formations.

## 5.1 OPOK! Principle

First major principle in Kannudi implementation of *One Phoneme One Key* (OPOK!) Here, keys are assigned to phonemes, not graphemes; and further, a one-one correspondence exists between a key and a phoneme. Key assignments adhere to the standard specified by Government of Karnataka.

Figure 1. Keyboard Layout – non-Shift State

Figure 2. Keyboard Layout – Shift State

## 5.2 Default Vowel – *null* ್ or *a* ಅ

A pure phonetic method of input would be to type every vowel and consonant in the order pronounced. This assumes no vowel inherently attached to a consonant in the alphabet. Thus, with OPOK! in force, ಕ್ needs only one keystroke but each *kāguṇita ka kā ki... kau* ಕ, ಕಾ, ಕಿ, ...ಕೌ requires 2 keystrokes.

As can be seen from the Table 1, majority of the syllables are *mūla akṣara*s, i.e., independent vowels and consonants with vowel *a* ಅ. Hence, in the alphabet, the graphemes for consonants have been designed with an assumed or default vowel (*ūhita svara*) *a* ಅ.

*Ottu* is the secondary form of a consonant that appears in a conjunct (*sanyuktākṣara*). An *otttu* is produced when there is no vowel between the two consonants, indicated by *null* vowel or virama ್.

If the default vowel is virama ್, then one simply types consonant keys in succession. However, if the default vowel *a* ಅ, then the *null* vowel must be typed in with an additional keystroke. For example, to type ಕ್ತ three keys need to be pressed as shown in Table 2.

It adds an "extra" key, thus significantly negating the savings provided by the default vowel. Even then, having *a* ಅ as the default vowel saves 24% of keystrokes (42-18=24%) compared to having no default vowel (or assuming *null* ್ as the default vowel). However, one may find this somewhat unnatural and not a pure phonetic method, and experience a loss in the speed of typing.

| Default Vowel | *a* ಅ | | | *null* ್ | | |
|---|---|---|---|---|---|---|
| Key pressed | k | f | t | k | t | a |
| Result | ಕ | ಕ್ | ಕ್ತ | ಕ್ | ಕ್ತ್ | ಕ್ತ |

Table 2: Typing an *ottakshara* conjunct

As such, Kannudi provides a choice of default vowels *a* ಅ and *null* ್. A user can choose one as a preference or switch between the two as suitable.

## 5.3 OHOK! Principle

OHOK! uses pressure as input to produce secondary forms of consonants (*ottu*). It is to apply pressure or press and hold a key to produce an *ottu*.

In case of mechanical keyboards, which are not pressure sensitive, OHOK!, which can be thought of as ***Otti Hidi Ottu Kodu!*** ಒತ್ತಿ ಹಿಡಿ ಒತ್ತು ಕೊಡು!, is simply time based, that is, dependent on how long the key is held pressed. Though keyboard timings, touch, and pressure sensitivities can be optimized to speed OHOK! we understand that it may be beyond the normal capabilities of a user.

In case of pressure sensitive (mobile) devices, OHOK! is called *Ottu Hāku Ottu Koḍu!* ಒತ್ತು ಹಾಕು ಒತ್ತು ಕೊಡು! (apply pressure and give *ottu*). This offers greater potential for savings in time as well as key strokes.

OHOK! can have three modes. Here OHOK! of a key, namely applying pressure (or holding pressed) is denoted by superscript $^+$ sign.

1. *Sva-ottu* **Mode** (SO): This is Self-Ottu, also known as *Dvitva* where a consonant gets its own secondary form (*ottu*) attached. For example, $k^+$ will produce ಕ್ಕ. And $kn^+w$, equivalent to typing *knfnw*, produces ಕನ್ನಡ.
2. *Kaṇḍante Ottu* **Mode** (KO): This is visual mode where the input follows the written order – holding a consonant will add its *ottu* to preceding consonant. For example, the key sequence $st^+r^+I$ (= *sftfrI*) produces ಸ್ತ್ರೀ.
3. *Andante Ottu* **Mode** (AO): This is "as you say" or phonetic mode where the input follows the order of pronunciation. Pressing and holding (ಒತ್ತಿ ಹಿಡಿ) of a consonant key will prepare it for an *ottu* by adding the *null* vowel (virama). This agrees with the order of pronunciation as the *ottu* (accent) is on this consonant and the next consonant is turned into an *ottu*. Here $k^+$ is equivalent to *kf*. To produce ಸ್ತ್ರೀ enter $s^+t^+rI$. In essence this is an alternative to *f* key in normal mode.

| **Default Vowel = *a* ಅ** | | | | | | | | | | |
|---|---|---|---|---|---|---|---|---|---|---|
| **Syllable Type** | | **Mode** | | | | | | | | |
| | | Normal | | | OHOK! SO *Dvitva* | | | OHOK! KO *Kandante* | | OHOK! AO *Andante* | |
| **Self-conjunct (dvitva)** | Keys Typed | k | f | k | $k^+$ | | | k | $k^+$ | $k^+$ | k |
| | Display | ಕ | ಕ್ | ಕ್ಕ | ಕ್ಕ | | | ಕ | ಕ್ಕ | ಕ್ | ಕ್ಕ |
| **Non-self-conjunct** | Keys Typed | g | f | r | g | f | r | g | $r^+$ | $g^+$ | r |
| | Display | ಗ | ಗ್ | ಗ್ರ | ಗ | ಗ್ | ಗ್ರ | ಗ | ಗ್ರ | ಗ್ | ಗ್ರ |
| **End-Virama** | Keys Typed | n | f | | n | f | | n | f | $n^+$ | |
| | Display | ನ | ನ್ | | ನ | ನ್ | | ನ | ನ್ | ನ್ | |
| **V[MH]? + self-conjunct** | Keys Typed | ak | f | k | $ak^+$ | | | ak | $k^+$ | $ak^+$ | k |
| | Display | ಅಕ | ಅಕ್ | ಅಕ್ಕ | ಅಕ್ಕ | | | ಅಕ | ಅಕ್ಕ | ಅಕ್ | ಅಕ್ಕ |
| **Keys saved per 1000 syllables** | | **0** | | | **174** | | | **196** | | **193** | |

Table 3: Key savings with *a*-default

| **Syllable Type** | | **Default Vowel = null** ಀ | | | | | | |
|---|---|---|---|---|---|---|---|---|
| | | **Mode** | | | | | | |
| | | Normal | | OHOK! *Dvitva* | | OHOK! KO *Kandante* | | OHOK! AO *Andante* | |
| **Self-conjunct (dvitva)** | Keys Typed | k | k | k$^+$ | | k | k$^+$ | k$^+$ | k |
| | Display | ಕ್ | ಕ್ಕ್ | ಕ್ಕ್ | | ಕ್ | ಕ್ಕ್ | ಕ್ | ಕ್ಕ್ |
| **Non-self-conjunct** | Keys Typed | g | r | g | r | g | r$^+$ | g$^+$ | r |
| | Display | ಗ್ | ್ರ | ಗ್ | ್ರ | ಗ್ | ್ರ | ಗ್ | ್ರ |
| **Virama** | Keys Typed | n | | n | | | | n$^+$ | |
| | Display | ನ್ | | ನ್ | | ✕ | | ನ್ | |
| **V[MH]? + self-conjunct** | Keys Typed | ak | k | ak$^+$ | | ak$^+$ | | ak$^+$ | k |
| | Display | ಅಕ್ | ಅಕ್ಕ್ | ಅಕ್ಕ್ | | ಅಕ್ಕ್ | | ಅಕ್ | ಅಕ್ಕ್ |
| **keys saved per 1000 syllables** | | **0** | | **87** | | **11** | | **0** | |

Table 4: Key Savings with *null*-default

## 5.4 Key Savings

Table 3 shows the key savings, with *a*-default, for various conjunct syllable types for the three modes of OHOK!. For example, a *dvitva* requiring 3 keys normally, can be produced with only 1 key in *dvitva* mode (SO).

Similarly, Table 4 lists the key savings with *null*-default.

These tables also show the keys saved per 1000 syllables in a typical document, based on the frequency data in Table 1. Thus, under a-default, key savings of 174 can be achieved in OHOK! *Dvitva* mode (SO).

It may be noted that, most key savings occur in *a*-default mode while the differences among the three OHOK! modes remain insignificant.

**Another scheme**: when a key is pressed and held next key becomes an *ottu*. e.g press-holding a *k* then typing *r* will produce ಕ್ರ in *a-default*-mode. This scheme has physical limitations due to positions of keys being fixed on a keyboard. It becomes necessary to cross hands or fingers or quickly decide which hand to use for a key. This is contrary to the trained typist who expects to blindly use the same finger at the same physical location for a given key. It can be physically impossible or totally confusing. Hence this scheme is not implemented in Kannudi.

## 6 Rules of Convenience

Apart from introducing the novel input method, Kannudi implements several user-friendly features that are simply matter of convenience or eliminate or reduce errors. A few such features are described here.

### 6.1 Backspace ←BS

During normal course of tying, a user may have typed a key in error or pressed an adjacent key. When the mistake is realized, the normal action is to press *backspace* key ←BS. This usually deletes the previous syllable entirely. If in midst of a multi-phoneme syllable, all the effort is lost, when the user desires to undo just last entry or the mistake just made. Hence, Kannudi recognizes 4 types of deletions:

1. Phoneme Delete: delete the phoneme immediately left of the cursor, assigned to ←BS.
2. Character Delete: delete the (unicode) character immediately left of the cursor, assigned to *alt←BS*,
3. Syllable Delete: delete the syllable immediately left of the cursor, assigned to *shift←BS*,
4. Word Delete: delete the word immediately left of the cursor, assigned to *ctrl←BS*.

## 6.2 Shunyification

*Sonne* or *śūn'ya* is used in writing to represent an *anunāsika* before a consonant as in ಅಂಕ, though it is not incorrect to write ಅಙ್ಕ. Thus, s*onne* before a classified consonant is pronounced as the *anunāsika* of the same class. The process of automatic conversion of *anunāsika* before a consonant, classified or non-classified, to a *sonne* is called *shunyification*.

For the sake of convenience, and only *n/m* keys are considered for *Śūn'yīfication* in this implementation. *Śūn'yīfication* makes the input entry flexible by allowing both *n* and *m* to be automatically *Śūn'yīfied*.

*Śūn'yīfication* is straightforward in case of a classified consonant but can be ambiguous in case of a non-classified consonant and exceptions occur.

In most cases when *sonne* precedes a non-classified consonant *(avargīya vyañjana)*, *Śūn'yīfication* allows the entry to be phonetic corresponding to how most people pronounce. For example, ಸಂಶಯ is pronounced as ಸಮ್ಶಯ *samśaya* and ಸಂಸ್ಕೃತ as sanskṛta or samskṛta by many, *albeit* all incorrectly.

*Śūn'yīfication* does not save any keystrokes; But does not require one to switch the flow between *andante* (as pronounced) and *kaṇḍante* (as seen). This is especially convenient for those who are mentally "spelling" the Kannada words in roman script as they type. And there are many such casual users.

## 6.3 Arkification

Though *arkāvottu* ೯ is used mostly in place of ರ್, there are a few situations where it is not desirable as in ಡ್ಯಾಂಕ, ಸರ್ನೆ. Kannudi automatically uses the correct form (*arkifies*) in such cases allowing the user to type normally without hindrance.

# 7 Error Prevention Rules

Certain domain knowledge of the language can be used to prevent typos and warn the user. A few examples are described below.

## 7.1 Non-initial Vowel

Kannada allows a standalone vowel only in beginning of word. Kannudi prevents such typing.

## 7.2 Aspirated Ottu

It is not possible to pronounce an aspirated consonant (*mahāprāṇa*) when followed by another aspirated consonant. As such Kannada does not allow a *mahāprāṇa ottu* to another *mahāprāṇa*. However, certain words can be exceptions due to common usage, *e.g.*, ವಿಠ್ಠಲ *viṭhṭhala*.

# 8 Exception Handling

Convenience and error prevention rules may not be perfect, and exceptions can be found as mentioned earlier. Hence it is necessary to ensure that there is a mechanism to override the normal behavior. Two ways to override are a) using ZWJ/ZWNJ characters and b) inputting the phonemes with a space character in between and then remove it.

# 9 Conclusions

Here we have introduced an input method called OHOK! with three possible modes, namey, *sva-ottu* (self-conjunct)*, kaṇḍante* (as seen), and *andante* (as pronounced/said*)*. It may be noted that *kaṇḍante* mode (KO) does not follow the phonemic order. However, this mode may work well for those who are inclined to visualize as they type rather than vocalizing the sounds.

OHOK! will work very well on mobile or any device where input pressure can be sensed where OHOK! can be called O*ttu Hāku Ottu Koḍu!* (Apply Pressure and Give *Ottu*) when implemented on pressure sensitive devices. it where can be a real time saver. On mechanical keyboards, it saves keystrokes though any time saved is dependent on keyboard settings.

We have showed that domain knowledge can be used to improve user friendliness. Several convenience and error minimization rules such as *Śūn'yīfication* and *arkāvottu* are described. Four types of deletions, namely phoneme, character, syllable, and word delete are identified and assigned to backspace key. Thus, domain knowledge is shown to be necessary and helpful to enhance user friendliness as wells as input flow and speed.

Further, one may consider incorporating these rules into open type font tables as attached language specific resources, and eliminate the need for a separate editor application.

## Terminology

| | |
|---|---|
| *andante* | As pronounced/said |
| *anunāsika* | Nasal consonant, fifth member of each of the 5 classes of consonants, ṅ, ñ, ṇ, n, and m |
| *arkāvottu* | *rēpha,* symbol ೯ for sound r |
| *arkīfication* | Automatic conversion of r to arkāvottu ೯ |
| *dvitva* | self-conjunct |
| *kāguṇita* | Consonant + Vowel, C+V |
| *kaṇḍante* | as seen (as written) |
| *mahāprāṇa* | Aspirated consonant |
| *mahāprāṇa* | Aspirated consonant |
| *mūla akṣara* | vowel a or C+a |
| *null vowel* | Represnted with *virāma* ್ |
| *ottakshara* | Syllable with *ottu,* Conjunct |
| *Otti hiḍi ottu koḍu!* | Hold pressed give *ottu* |
| *ottu* | 1. secondary form of a consonant; 2. accent |
| *ottu Hāku Ottu Koḍu!* | apply pressure give *ottu* |
| *rēpha* | Symbol ೯ for sound r |
| *sanyuktākṣara* | conjunct |
| *sonne* | śūn'ya, anusvara, symbol ಂ |
| *śūn'ya* | sonne, anusvara, symbol ಂ |
| *Śūn'yīfication* | Automatic conversion of *anunāsika* to *śūn'ya* |
| *Sva-ottu* | self-conjunct |
| *ūhita svara* | Default/Presumed vowel |
| *virāma* | Null vowel, soundless vowel symbol ್ |